\documentclass{article}

\usepackage{arxiv}

\usepackage[utf8]{inputenc} 
\usepackage[T1]{fontenc}    
\usepackage{hyperref}       
\usepackage{url}            
\usepackage{booktabs}       
\usepackage{amsfonts}       
\usepackage{nicefrac}       
\usepackage{microtype}      
\usepackage{cleveref}       
\usepackage{graphicx}
\usepackage[numbers,square]{natbib}
\usepackage{doi}
\usepackage{multirow}
\usepackage[table,xcdraw]{xcolor}
\usepackage{float}
\usepackage{longtable}

\title{ATRAF-driven IMRaD Methodology:\\\large Tradeoff and Risk Analysis of Software Architectures\\ Across Abstraction Levels}

\date{}

\usepackage{authblk}

\newbox{\orcid}\sbox{\orcid}{\includegraphics[scale=0.06]{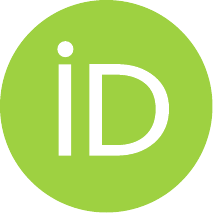}}

\author[1,2,3]{%
	\href{https://orcid.org/0009-0005-8915-7905}{\usebox{\orcid}\hspace{1mm}Amine Ben~Hassouna\thanks{\texttt{amine.benhassouna@medtech.tn, amine.benhassouna@dracodes.com (Corresponding author)}}}%
}

\affil[1]{Mediterranean Institute of Technology, South Mediterranean University, Tunis, Tunisia}
\affil[2]{Dracodes, Tunis, Tunisia}
\affil[3]{ENSI - École Nationale des Sciences de l'Informatique, Manouba, Tunisia}

\renewcommand{\headeright}{Preprint}
\renewcommand{\undertitle}{Preprint}
\renewcommand{\shorttitle}{ATRAF-driven IMRaD Methodology}

\hypersetup{
  pdftitle={ATRAF-driven IMRaD Methodology},
  pdfsubject={cs.SE},
  pdfauthor={Amine Ben Hassouna},
  pdfkeywords={ATRAF, IMRaD Methodology, Software architecture, Architectural frameworks, Tradeoff analysis, Quality attributes, Architecture evaluation},
}

\begin{document}
\maketitle

\begin{abstract}
Software architecture research relies on key architectural artifacts---Software Architectures, Reference Architectures, and Architectural Frameworks---that underpin the design and analysis of complex systems. Evaluating these artifacts is essential to assess tradeoffs and risks affecting quality attributes such as performance, modifiability, and security. Although methodologies like the Architecture Tradeoff Analysis Method (ATAM) support software architecture evaluation, their industrial focus misaligns with the IMRaD (Introduction, Methods, Results, Discussion) format prevalent in academic research, impeding transparency and reproducibility. Our prior work introduced the Architecture Tradeoff and Risk Analysis Framework (ATRAF), extending ATAM through three methods---ATRAM, RATRAM, and AFTRAM, addressing all abstraction levels, using a unified, iterative four-phase spiral model. These phases---Scenario and Requirements Gathering, Architectural Views and Scenario Realization, Attribute-Specific Analyses, and Sensitivity, Tradeoff, and Risk Analysis---ensure traceability and coherence. This paper presents the ATRAF-driven IMRaD Methodology, a concise method to align ATRAF's phases with IMRaD sections.  This methodology enhances the rigor, transparency, and accessibility of software architecture research, enabling systematic reporting of complex evaluations.
\end{abstract}

\keywords{ATRAF \and IMRaD Methodology \and Software Architecture \and Reference Architecture \and Architectural Framework \and Architecture Evaluation \and Tradeoff Analysis \and Quality Attributes \and Risk Analysis}


\section{Introduction}
\label{sec:introduction}

In the realm of software engineering, architectural artifacts---including Software Architectures (SAs), Reference Architectures (RAs), and Architectural Frameworks (AFs)---serve as foundational constructs that guide system design, reuse, and evaluation across varying abstraction levels. These artifacts significantly shape quality attributes, including performance, modifiability, and security. Rigorously evaluating them is essential to identify tradeoffs and risks that could undermine system quality. However, existing evaluation methods, such as the Architecture Tradeoff Analysis Method (ATAM)~\cite{kazman1998atam, kazman2000atam}, evaluate only software architectures, while ATAM/R~\cite{angelov2014atamr} extends solely to reference architectures. Neither method addresses architectural frameworks. Moreover, these industrial methods misalign with the IMRaD format---standard in academic research for its Introduction, Methods, Results, and Discussion structure---hindering transparency and reproducibility.

To bridge this gap, our prior work introduced the Architecture Tradeoff and Risk Analysis Framework (ATRAF)~\cite{benhassouna2025atraf}, extending ATAM with three methods: Architecture Tradeoff and Risk Analysis Method (ATRAM) for Software Architectures, Reference Architecture Tradeoff and Risk Analysis Method (RATRAM) for Reference Architectures, and Architectural Framework Tradeoff and Risk Analysis Method (AFTRAM) for Architectural Frameworks. ATRAF employs a scenario-driven, iterative four-phase spiral model to evaluate artifacts across abstraction levels, addressing the limitations of ATAM and ATAM/R for abstract artifacts. Using ATRAF in academic research, however, requires a structured approach to align its iterative process and diverse artifacts with IMRaD's linear narrative, enhancing clarity and reproducibility in reporting.

This study introduces the ATRAF-driven IMRaD Methodology to integrate ATRAF's evaluation methods into IMRaD-structured research papers. This methodology aligns ATRAF's four phases---Scenario and Requirements Gathering, Architectural Views and Scenario Realization, Attribute-Specific Analyses, and Sensitivity, Tradeoff, and Risk Analysis---with IMRaD sections, ensuring a coherent, traceable evaluation. It resolves the misalignment between industrial evaluation methods and academic reporting standards, improving the rigor, transparency, and accessibility of software architecture research.

The remainder of this paper is organized as follows: Section~\ref{sec:background} provides background on ATRAF and IMRaD, Section~\ref{sec:methodology} details the ATRAF-driven IMRaD Methodology, Section~\ref{sec:case-studies} presents case studies of research papers applying the methodology to illustrative examples from the ATRAF paper, and Section~\ref{sec:conclusion} summarizes findings and discusses future research directions.


\section{Background}
\label{sec:background}

Software engineering relies on architectural artifacts---Software Architectures (SAs), Reference Architectures (RAs), and Architectural Frameworks (AFs)---to shape the design and evaluation of systems, differing in abstraction and scope. SAs offer concrete blueprints, detailing system structure, components, and interactions to meet functional and quality requirements. RAs, at a higher abstraction, provide reusable templates that embody best practices and design patterns, enabling development of domain-specific architectures. AFs, operating at the highest abstraction level, define abstract templates and meta-level methodologies to support SA and RA development across domains, enabling instantiation into concrete architectures while ensuring flexibility through process-oriented guidance.

ATRAF provides a unified, scenario-driven framework to evaluate SAs, RAs, and AFs, addressing their unique tradeoffs and risks with specialized methods~\cite{benhassouna2025atraf}. Extending ATAM~\cite{kazman1998atam, kazman2000atam}, it offers three methods---ATRAM, RATRAM, and AFTRAM---each tailored to specific artifacts but following a shared four-phase spiral process. ATRAM evaluates system-specific SAs, iteratively assessing quality attributes like performance and security. RATRAM supports RAs, focusing on domain-specific reuse and variability through generalized scenarios and instantiated architectures. AFTRAM adapts RATRAM for AFs, targeting meta-level extensibility and lifecycle support across system families.

The four-phase spiral process comprises:
\begin{enumerate}
    \item \textbf{Phase I, Scenario and Requirements Gathering:} This phase elicits functional scenarios for ATRAM, domain-aligned scenarios for RATRAM, and multi-context scenarios for AFTRAM, alongside corresponding system-, domain-, or framework-level requirements.
    \item \textbf{Phase II, Architectural Views and Scenario Realization:} This phase constructs architectural views---structural, interaction, behavioral, and deployment for ATRAM; augmented with variability for RATRAM; incorporating process-oriented views and extensibility for AFTRAM---and maps scenarios to architectural elements, including instantiation for RATRAM and AFTRAM.
    \item \textbf{Phase III, Attribute-Specific Analyses:} This phase evaluates quality attributes independently using established views and mappings.
    \item \textbf{Phase IV, Sensitivity, Tradeoff, and Risk Analysis:} This phase identifies sensitivity points, tradeoffs, and risks, consolidating findings to support iterative refinement.
\end{enumerate}

ATRAF's process ensures consistency across artifact types, with RATRAM's variability and AFTRAM's extensibility views addressing abstraction demands~\cite{benhassouna2025atraf}. However, its iterative spiral process contrasts with IMRaD's linear narrative, the standard for academic research. While IMRaD fosters clarity and reproducibility through its sequential structure, it struggles to capture ATRAF's iterative cycle. Section~\ref{sec:methodology} proposes a method to align ATRAF's process with IMRaD's organization, enhancing rigor and accessibility in reporting architectural evaluations.


\section{Methodology}
\label{sec:methodology}

This methodology integrates ATRAF's evaluation methods---ATRAM, RATRAM, and AFTRAM---into IMRaD-structured research papers. This section adapts ATRAF's iterative process to IMRaD's linear narrative and details its phases' mapping to IMRaD sections.

\begin{figure}[H]
    \centering
    \includegraphics[width=\textwidth]{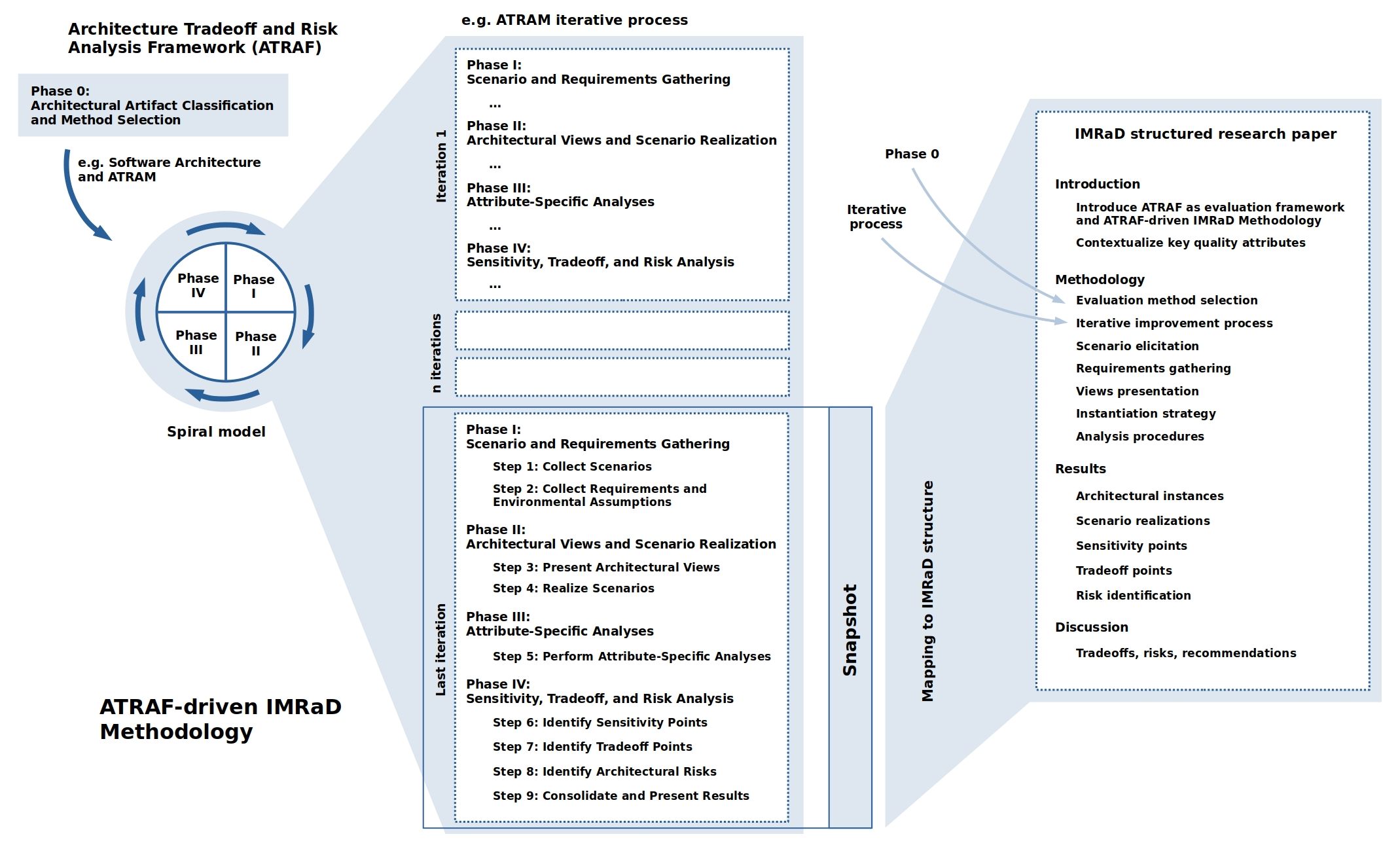}
    \caption{Overview of the ATRAF-driven Methodology}
    \label{fig:atraf-driven-imrad-methodology}
\end{figure}

\subsection{Integrating ATRAF's Spiral Process into IMRaD}
\label{sec:adapting-spiral-process}

The ATRAF-driven IMRaD Methodology aligns ATRAF's iterative process with IMRaD's linear narrative by presenting the research paper as a final snapshot of the iterative process undertaken by the researcher to develop and evaluate the artifact. This snapshot approach ensures a coherent and logical presentation, maintaining traceability to insights from ATRAF's iterative cycles---scenario realizations, attribute analyses, tradeoff identifications. To implement this, the methodology recommends researchers describe the ATRAF process in the Methodology section, summarizing key refinements---risk mitigations, for example---that refined the artifact or evaluation process while upholding IMRaD's linear narrative. Figure~\ref{fig:atraf-driven-imrad-methodology} depicts how each phase aligns with IMRaD sections, promoting transparency and highlighting the methodology's iterative rigor.

\subsection{Mapping ATRAF to IMRaD}
\label{sec:mapping-atraf-to-imrad}

This methodology assigns ATRAF's phases to IMRaD sections, ensuring traceability and flexibility for ATRAM, RATRAM, and AFTRAM. It incorporates essential artifacts---utility trees, tradeoff matrices---into the paper's main body, while non-essential artifacts---stakeholder maps, for example---may appear in the Appendix, as supplementary material, or be omitted with justification---e.g., resource constraints. Such omissions must be documented in the Methodology section to maintain transparency.

\subsubsection{Recommended Introduction Structure}
\label{sec:introduction-structure}

For papers addressing software architecture, reference architecture, or architectural frameworks, the ATRAF-driven IMRaD Methodology refines the Introduction's structure to present research rigorously. A standard Introduction establishes the domain context, identifies challenges, articulates a problem statement, positions the solution within the literature, outlines evaluation methods, and summarizes contributions. To enhance this, the methodology augments the structure by explicitly introducing ATRAF as the evaluation framework for creating and assessing the architectural artifact. It recommends researchers reference this methodology to justify integrating ATRAF's scenario-driven process into IMRaD's linear format. Additionally, it advises researchers to contextualize key quality attributes---performance, security, modularity---in their Introduction, linking them to scenarios and requirements to clarify the artifact's design rationale. These enhancements foster a transparent, reproducible Introduction aligned with ATRAF's iterative, scenario-based process.

\subsubsection{Phase Mapping}
\label{sec:atraf-phases-mapping}

The ATRAF-driven IMRaD Methodology initiates ATRAF's evaluation process with Phase 0, Architectural Artifact Classification and Method Selection, identifying the artifact's class to select the evaluation method---ATRAM for SAs, RATRAM for RAs, or AFTRAM for AFs. This classification, per the ATRAF paper~\cite{benhassouna2025atraf}, applies the Goals-Inputs-Outcomes (GIO) model and flexible principles to categorize artifacts as system-specific, domain-specific, framework-based, or hybrid---e.g., SA with RA traits or RA with AF traits. The methodology recommends researchers document this classification and method selection in the Methodology section before proceeding to subsequent phases, ensuring transparency.

Building on Phase 0's classification, the ATRAF-driven IMRaD Methodology maps ATRAF's core phases---Scenario and Requirements Gathering, Architectural Views and Scenario Realization, Attribute-Specific Analyses, and Sensitivity, Tradeoff, and Risk Analysis---to IMRaD sections. It recommends researchers document these mappings in the Methodology section for transparency. For Phase I, scenario elicitation and requirements gathering should be presented in the Methodology section. For Phase II, architectural views and instantiation should be outlined in the Methodology section, while scenario mappings, architectural instantiations, and meta-scenario realization should be reported in the Results section. For Phase III, analysis procedures should be specified in the Methodology section, with outcomes documented in the Results section. For Phase IV, sensitivity points, tradeoff points, and architectural risks should be presented in the Results section, with interpretations and recommendations elaborated in the Discussion section. Table~\ref{tab:atraf-phase-mapping} details these mappings, clarifying variations across ATRAM, RATRAM, and AFTRAM.

\begin{table}[h]
    \caption{Mapping of ATRAF Phases and Steps to IMRaD Sections}
    \label{tab:atraf-phase-mapping}

    \centering
    {\scriptsize 
    \begin{tabular}{|c|p{0.2\textwidth}|p{0.2\textwidth}|p{0.2\textwidth}|p{0.2\textwidth}|}
        \hline
        \textbf{Phase} & \textbf{ATRAM steps} & \textbf{RATRAM steps} & \textbf{AFTRAM steps} & \textbf{IMRaD Mapping} \\
        \hline
        \textbf{Phase 0} & \multicolumn{3}{c|}{Identify architectural artifact and select method} & \textbf{Methods} (method selection) \\
        \hline
        \multirow{2}{*}{\textbf{Phase I}} & Collect Scenarios & Collect Domain-Aligned Scenarios & Collect Multi-Context Scenarios & \textbf{Methods} (scenario elicitation) \\
        \cline{2-5}
        & Collect Requirements and Environmental Assumptions & Collect Requirements and Environmental Assumptions & Collect Requirements and Environmental Assumptions & \textbf{Methods} (requirements gathering) \\
        \hline
        \multirow{3}{*}{\textbf{Phase II}} & Present Architectural Views & Present Reference Architecture Views & Present Architectural Framework Views and Processes & \textbf{Methods} (views presentation) \\
        \cline{2-5}
        &  & Instantiate and Describe Architectural Instances & Instantiate and Describe Architectural Instances  & \textbf{Methods} (instantiation strategy) \newline \textbf{Results} (architectural instances) \\
        \cline{2-5}
        & Realize Scenarios & Map and Realize Scenarios through Architecture Instances & Map and Realize Scenarios through Architecture Instances & \textbf{Results} (scenario realizations) \\
        \cline{2-5}
        &  &  & Map and Realize Framework-Level Requirements through Meta-Scenario Simulation & \textbf{Results} (meta-scenario realizations) \\
        \hline
        \textbf{Phase III} & Perform Attribute-Specific Analyses & Perform Attribute-Specific Analyses & Perform Attribute-Specific Analyses & \textbf{Methods} (analysis procedures) \newline \textbf{Results} (attribute evaluations) \\
        \hline
        \multirow{4}{*}{\textbf{Phase IV}} & Identify Sensitivity Points & Identify Sensitivity Points & Identify Sensitivity Points & \textbf{Results} (sensitivity points) \\
        \cline{2-5}
        & Identify Tradeoff Points & Identify Tradeoff Points & Identify Tradeoff Points & \textbf{Results} (tradeoff points) \\
        \cline{2-5}
        & Identify Architectural Risks & Identify Architectural Risks & Identify Architectural Risks & \textbf{Results} (risk identification) \\
        \cline{2-5}
        & Consolidate and Present Results & Consolidate and Present Results & Consolidate and Present Results & \textbf{Discussion} (tradeoffs, risks, recommendations) \\
        \hline
    \end{tabular}
    } 
\end{table}

This mapping strategy ensures that the evaluation process is traceable, accommodates the differences in abstraction levels across architectural artifacts, and adheres to the structural constraints of the IMRaD format.

\subsection{Adoption Levels}
\label{sec:adoption-levels}

To accommodate diverse research objectives and resource constraints, the ATRAF-driven IMRaD Methodology offers Light and Full adoption levels. These levels vary in evaluation depth, artifact presentation, and tradeoff and risk analysis granularity, allowing researchers to adapt the methodology effectively.

For ATRAF adoption, the ATRAF-driven IMRaD Methodology recommends researchers adapt artifacts and stakeholder engagement to balance transparency, rigor, and resource constraints. Essential artifacts---utility trees, tradeoff matrices---should appear in the Results or Discussion sections to ensure clarity. Non-essential artifacts---stakeholder maps, for example---may appear in the Appendix, as supplementary material, or be omitted, with justifications, such as reliance on literature-based scenarios, documented in the Methodology section. Similarly, stakeholder engagement---workshops for scenario elicitation, for instance---should be adapted or omitted based on constraints, with deviations justified in the Methodology section to maintain rigor.

\subsubsection{Light Adoption}
\label{sec:light-adoption}

Light adoption streamlines the ATRAF-driven IMRaD Methodology for studies prioritizing the architectural artifact over exhaustive tradeoff and risk evaluation. It suits limited-scope or resource-constrained research, emphasizing the artifact's design and key outcomes. Scenario elicitation selectively targets critical quality attributes, while utility trees may be simplified as narrative summaries rather than formal diagrams or woven into the text. Attribute analyses focus on key attributes, and non-essential artifacts---stakeholder maps, requirement traceability matrices, scenario logs---may be omitted or briefly summarized in the Methodology section to avoid overwhelming the reader. Omissions, due to resource constraints, should be justified in the Methodology section---e.g., citing scenario log volume---while artifacts may appear in supplementary material or the Appendix with justification. Tradeoff and risk analyses, kept minimal, integrate into the narrative, aligning with a concise IMRaD structure to balance rigor and artifact focus.

\subsubsection{Full Adoption}
\label{sec:full-adoption}

In contrast, Full adoption maximizes the ATRAF-driven IMRaD Methodology for research requiring comprehensive evaluations, with tradeoff and risk analyses central to the contribution. It suits complex or high-stakes research prioritizing transparency and reproducibility. Full adoption applies ATRAF rigorously, with thorough scenario elicitation, tradeoff and risk analyses, and detailed artifacts. It executes all four phases, mapping outcomes to IMRaD sections per Table~\ref{tab:atraf-phase-mapping}. Essential artifacts---utility trees, scenario mappings, tradeoff matrices---appear in the main body, with detailed figures or tables, while other artifacts may appear in the Appendix or as supplementary materials. It comprehensively analyzes multiple quality attributes, documents sensitivity points, tradeoff points, and risks in the Results section, and explores implications in the Discussion section. This level ensures comprehensive evaluation, transparency, and reproducibility, distinguishing it from Light adoption's selective approach.


\section{Case Studies and Discussion}
\label{sec:case-studies}

In future versions of this paper, case studies will demonstrate the ATRAF-driven IMRaD Methodology by producing research papers that create and evaluate architectures, unlike their industry-oriented presentation in the ATRAF paper~\cite{benhassouna2025atraf}. Originally, the Remote Temperature Sensor Architecture (RTSA), Remote Temperature System Reference Architecture (RTSRA), and Remote Monitoring Architectural Framework (RMAF) were introduced using ATRAF's methods---ATRAM, RATRAM, and AFTRAM---with a focus on practical application. By contrast, these future case studies will reframe RTSA, RTSRA, and RMAF as academic research papers, for example, creating and evaluating RTSA using ATRAM, structured by the ATRAF-driven IMRaD Methodology. Employing ATRAM for RTSA, RATRAM for RTSRA, and AFTRAM for RMAF, each study will produce complete IMRaD-formatted papers, showcasing the methodology's alignment of ATRAF's evaluation process with academic standards. This approach will ensure transparent, reproducible research, guiding researchers and practitioners in applying the methodology.


\section{Conclusion and Future Work}
\label{sec:conclusion}

This paper presents the ATRAF-driven IMRaD Methodology, a transformative approach that enables researchers to apply the Architecture Tradeoff and Risk Analysis Framework (ATRAF) within IMRaD-structured research papers. By mapping ATRAF's evaluation methods---ATRAM, RATRAM, and AFTRAM---to IMRaD sections, the methodology addresses the misalignment between industrial architectural evaluation practices and academic reporting standards, as identified in prior methods like ATAM~\cite{kazman1998atam,kazman2000atam}. It ensures transparent, rigorous, and reproducible reporting of architectural evaluations, whether focused on Software Architectures, Reference Architectures, or Architectural Frameworks. The “final snapshot” approach effectively captures ATRAF's iterative process, while selective artifact inclusion streamlines complexity, making the methodology a robust tool for scholarly communication.

The significance of this methodology lies in its flexibility and accessibility, accommodating diverse research objectives and resource constraints. Light and Full adoption levels allow researchers to tailor evaluations, from streamlined artifact-focused studies to comprehensive tradeoff and risk analyses, as detailed in Section~\ref{sec:adoption-levels}. Flexible stakeholder engagement further enhances its applicability, enabling resource-constrained researchers to adapt processes like scenario elicitation while maintaining rigor through documented justifications. Planned case studies on the Remote Temperature Sensor Architecture (RTSA), Remote Temperature System Reference Architecture (RTSRA), and Remote Monitoring Architectural Framework (RMAF), as outlined in Section~\ref{sec:case-studies}, will reframe these architectures as academic research papers, demonstrating the methodology's versatility across abstraction levels. This approach not only bridges the academic-industry divide but also sets a new standard for transparent architectural evaluation in software engineering research.

Future research will refine the ATRAF-driven IMRaD Methodology to enhance its precision and scope. Potential directions include integrating the methodology with complementary evaluation techniques, such as model-based analysis, to enrich its analytical depth. Additional case studies beyond RTSA, RTSRA, and RMAF will validate the methodology's generalizability across emerging domains, such as cloud-native architectures or AI-driven systems. Collaborative efforts with industry practitioners will further explore stakeholder engagement strategies, ensuring the methodology's practicality in diverse contexts. These advancements will solidify the methodology's role as a cornerstone for rigorous, reproducible architectural evaluation in academic and applied settings.


\bibliographystyle{unsrtnat}
\bibliography{references}

\begin{thebibliography}{4}
\providecommand{\natexlab}[1]{#1}
\providecommand{\url}[1]{\texttt{#1}}
\expandafter\ifx\csname urlstyle\endcsname\relax
  \providecommand{\doi}[1]{doi: #1}\else
  \providecommand{\doi}{doi: \begingroup \urlstyle{rm}\Url}\fi

\bibitem[Kazman et~al.(1998)Kazman, Klein, Barbacci, Longstaff, Lipson, and Carriere]{kazman1998atam}
Rick Kazman, Mark Klein, Mario Barbacci, Tom Longstaff, Howard Lipson, and Jeromy Carriere.
\newblock The architecture tradeoff analysis method.
\newblock In \emph{Proceedings. Fourth IEEE International Conference on Engineering of Complex Computer Systems (Cat. No.98EX193)}, pages 68--78. IEEE, 1998.
\newblock \doi{10.1109/ICECCS.1998.706657}.

\bibitem[Kazman et~al.(2000)Kazman, Klein, and Clements]{kazman2000atam}
Rick Kazman, Mark Klein, and Paul Clements.
\newblock Atam: Method for architecture evaluation.
\newblock Technical report, Carnegie Mellon University, Software Engineering Institute, Pittsburgh, PA, 2000.

\bibitem[Angelov et~al.(2014)Angelov, Trienekens, and Grefen]{angelov2014atamr}
S.~Angelov, J.~Trienekens, and P.~Grefen.
\newblock Extending and adapting the architecture tradeoff analysis method for the evaluation of software reference architectures.
\newblock Technical Report 443, Eindhoven University of Technology, 2014.

\bibitem[{Ben Hassouna}(2025)]{benhassouna2025atraf}
Amine {Ben Hassouna}.
\newblock The architecture tradeoff and risk analysis framework (atraf): A unified approach for evaluating software architectures, reference architectures, and architectural frameworks, 2025.
\newblock URL \url{https://doi.org/10.48550/arXiv.2505.00688}.

\end{thebibliography}

\end{document}